\documentclass[]{aa}
\usepackage{natbib}
\usepackage{graphicx}
\usepackage{txfonts}
\def\EBVl{\mbox{E$^\infty_{\rm B-V}$}}
\def\EBV{\mbox{E$_{\rm B-V}$}}

\def\AV{\mbox{A$_{\rm V}$}}
\def\AVl{\mbox{A$^\infty_{\rm V}$}}

\def\HH{\mbox{H$_2$}}

\def\vlsr{\mbox{${\rm v}_{lsr}$}}
\def\sv{\mbox{${\sigma}_{\rm v}$}}

\def\nH2{\mbox{${\rm n}_\HH}$}

\def\NH2{{\rm N}({\rm H}_2)}
\def\pccc{~{\rm cm}^{-3}} 
 
\def\pcc {~{\rm cm}^{-2}}

\def\Tsub#1 {\mbox{${\rm T}_{\rm #1}$}}
\def\TK  {\Tsub K }

\def\arcsec{\mbox{$^{\prime\prime}$}} \def\arcmin{\mbox{$^{\prime}$}}

\def\degr{$^{\rm o}$}
\def\p{\mbox{$^+$}}
\def\cotw {\mbox{$^{12}$CO}}
\def\coth {\mbox{$^{13}$CO}}

\def\hcop{\mbox{{HCO\p}}}

\def\WCO{\mbox{W$_{\rm CO}$}}

\def\h13cop{\mbox{{H$^{13}$CO\p}}}

\def\c3h2{\mbox{C$_3$H$_2$}}
 
 \def\R0{R$_0$}

\def\ddeg{{}^\circ\kern-.1em}  
\def\Msun{{M_0}}  

\def\kms{\mbox{km\,s$^{-1}$}}

\def\E#1 {$10^{#1}$}
\def\E#1 {E{#1}}
\def\P#1,{$\nH2\TK~=~#1\times~10^4\pccc$~K}
\def\ec#1,#2,#3,{#1\,(#2)\E{#3}}

\def\zoph{$\zeta$ Oph}

\def\H3{\mbox{H$_3$}}

\def\RH2{\mbox{R$_{\rm G}$}}
\def\fH2{\mbox{f$_{\HH}$}}
\def\FH2{\mbox{F$_{\HH}$}}

\def\g13{\mbox{g$_{13}$}}

\newcommand{\supjerome}[1]{}

\title{Imaging galactic diffuse clouds: \\ 
 CO emission, reddening and turbulent flow in the gas around \zoph}
\author{H. S. Liszt\inst{1}, J. Pety\inst{2,3}, and K. Tachihara\inst{4}}
\institute{National Radio Astronomy Observatory,
           520 Edgemont Road,
           Charlottesville, VA,
           USA 22903-2475
\and       Institut de Radioastronomie Millim\'etrique,
           300 Rue de la Piscine,
           F-38406 Saint Martin d'H\`eres,
           France
\and       Obs. de Paris, 
           61 av. de l'Observatoire, 75014, Paris, 
           France
\and       National Astronomical Observatory of Japan, 
            2-21-1, Osawa, Mitaka, Tokyo 181-8588,
           Japan}

\begin{document}
\date{received \today}
\offprints{H. S. Liszt}
\mail{hliszt@nrao.edu}
%
\abstract
 {Most diffuse clouds are only known as kinematic features in absorption
   spectra, but those with appreciable \HH\ content may be visible in 
  the emission of such small molecules as CH, OH, and CO.}
{We interpret in greater detail the extensive observations of \cotw\ emission 
  from diffuse gas seen around the archetypical 
 line of sight to \zoph.}
 {The \cotw\ emission is imaged in position and position-velocity space,
 analyzed statistically, and then compared with maps of total reddening \EBVl\ and 
 with models of the C\p\ - CO 
  transition in \HH-bearing diffuse clouds.}
{ Around \zoph, \cotw\ emission appears in two distinct intervals of reddening
 centered near \EBVl\ $\approx$ 0.4 and 0.65 mag,  of which $\la$ 0.2 mag
  is  background material.  Within either interval, 
  the integrated \cotw\ intensity varies up to 6-12 K \kms, compared to 1.5 K \kms\
  toward \zoph.  Nearly 80\% of the individual profiles have velocity 
  dispersions \sv\ $< 0.6$ \kms, which are subsonic at the kinetic 
  temperature derived from \HH\ toward \zoph, 55 K.  Partly as a result, 
  \cotw\ emission exposes the internal, turbulent, supersonic 
  (1-3 \kms) gas flows with especial clarity in the cores of strong
  lines.  The flows are manifested as resolved velocity gradients in narrow, 
  subsonically-broadened line cores.}
{The scatter between N(CO) and \EBV\ in global, CO absorption line 
surveys toward bright stars is present in the gas seen around 
 \zoph, reflecting the  extreme sensitivity of N(\cotw) to ambient
 conditions.  The two-component nature of the 
  optical absorption toward \zoph\ is coincidental and  the star is 
  occulted by a single body of gas with a complex internal structure, 
  not by two distinct clouds. The very bright \cotw\ lines in diffuse 
 gas arise at N(\HH) $\approx 10^{21}\pcc$ in regions 
of modest density n(H) $\approx 200-500\pccc$ and somewhat more 
complete C\p-CO conversion.  Given the variety of structure in
the foreground gas, it is apparent that only large surveys of 
absorption sightlines can hope to capture the intrinsic
behavior of diffuse gas.}

\authorrunning{Liszt, Pety and Tachihara}
\titlerunning{CO, reddening and turbulence around \zoph}

\maketitle

\section{Introduction}

The line of sight to the nearby (140-160 pc) runaway O9.5 V star HD 149757, 
\zoph, has served as the 
archetype for detailed observational studies of the internal composition 
of diffuse (\AV $\la 1$ mag) clouds \citep{Her68,Mor75}, for optical/uv
detection of new molecules in diffuse gas \citep{MaiLak+01} and for theoretical
models of molecular gas in diffuse clouds \citep{BlaDal77,VanBla86,
vDiBla88,KopGer+96}.  The \HH-bearing portions of the gas occulting
 \zoph\ are dense enough to host appreciable column densities
of carbon monoxide, N(\cotw) $ \approx 2.4 \times 10^{15}~\pcc$ 
\citep{Mor75,WanPen+82, LamShe+94,SonWel+07}, and these
are readily detectable in mm-wave emission toward the star 
\citep{KnaJur76,Lis79,LanGla+87}.  They were very partially mapped in CO emission 
\citep{KopGer+96,Lis97}, as well as CH and OH \citep{Cru79,Lis97}.

\begin{figure*}
\includegraphics[height=19cm]{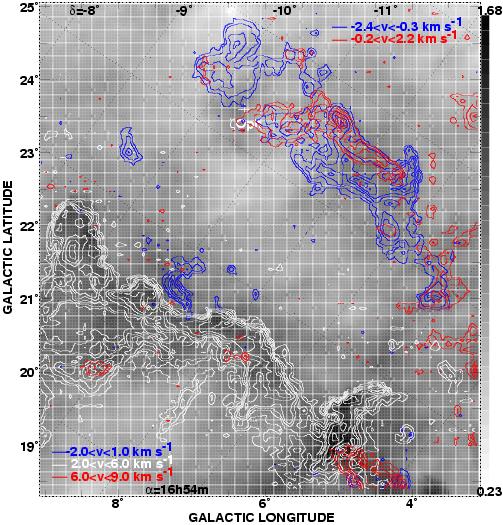}
\caption[]{Limiting reddening \citep{SchFin+98} and \cotw\ J=1-0 emission
\citep{TacAbe+00} in the vicinity of \zoph.  The \cotw\ emission contours
are shown at levels 1, 2, 4, 6, 8, 10, 12, 16, 20, 24 K \kms\ and 
have been calculated over different velocity ranges in the upper right and
lower left portions of the map, as indicated at the map corners.  The
dark cloud complex to the South of \zoph\ is known as L204  and 
the gas at the upper right, around \zoph, as L121.}
\end{figure*}

CO J=1-0 emission around \zoph\ was imaged in much more complete 
fashion by  \citet{TacAbe+00}, who focused their discussion on the properties 
of the nearby dark cloud complex L204 seen several degrees to the 
galactic South.  L204 is clearly outlined against the H$\alpha$ emission
from the ionized gas in the star's H II region \citep{GauMcC+01}.
In this work, the \cotw\ datacube from \citet{TacAbe+00} 
is employed to study the diffuse gas at \AV\ $\approx 1$ mag seen nearer the 
star.  We scrutinize the entire CO image of the absorption-line host whose 
overall properties have so often been inferred from one microscopic 
absorption sightline toward the star, and we inquire to what extent that 
line of sight faithfully represents the host gas.  Moreover, 
large-scale maps of reddening and extinction have become 
available at comparable resolution (though only along the entire line of sight, 
see \cite{SchFin+98} and \cite{DobUeh+05}), and we employ these to 
control against possible confusion between diffuse and darker 
sightlines, a source of concern given the strong CO lines we see.


The plan of this work is as follows.  Section 2 summarizes 
what is known observationally of the line of sight toward the star 
and describes the pre-existing H I, CO, \AVl\ and \EBVl\ datasets 
which are discussed here.  Section 3 discusses
the appearance of the sky around \zoph\ in terms of the statistics
of \cotw\ emission and reddening.  Section 4 discusses CO profiles
and linewidths in terms of  the turbulent flows which are prominent
in the emission profiles.  Section  5 discusses physical conditions
in the CO and \HH-bearing host gas, especially the regions of extremely
bright (11-12 K) CO emission.  Sect. 6 is a summary and Sect. 7 (available
online) discusses the relationship between the extinction and reddening
measurements over the \zoph\ field and presents some additional views
of the \cotw\ observations.

\section{Observations}

\subsection{Carbon monoxide}

The datacube of \cite{TacAbe+00} comprises nearly 11,000 spectra from
the NANTEN telescope with a beamwidth HPBW = 2.7\arcmin\ on a 
4\arcmin\ grid in galactic coordinates.  The spectra have 0.1 \kms\ 
resolution and the single-channel rms at this resolution, 0.5 K,  is 
relatively high compared to that in the small numbers of demonstration 
spectra typically shown in earlier work \citep{Lis97}.  Statisically
significant detections of the CO require \WCO\ $\ga 1$ \kms,
where \WCO\ is the integrated intensity

Toward the star, we show the profile of \cite{Lis97} from the then-NRAO Kitt 
Peak 12m telescope at 1\arcmin\ (HPBW) spatial resolution and 
0.12 \kms\ spectral resolution (see Sect. 3) .  To ensure that this
profile is compatible with those from NANTEN we recently used the ARO 12m
Kitt Peak telescope to re-observe several positions having comparatively 
strong emission in the NANTEN datacube.  The 12m spectra agree with the NANTEN
data to better than 5\%, a remarkable coincidence considering the 
difference in hardware and spatial resolution.

\subsection{H I}

\begin{figure}
\includegraphics[height=8.5cm]{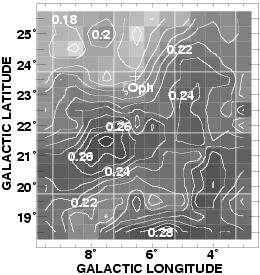}
\caption[]{Integrated H I line brightness from the LDSS survey of
\cite{HarBur97} at 35\arcmin\ resolution, converted to equivalent reddening
\EBV = N(H I)/$5.8\times 10^{21}\pcc$, N(H I) =$1.823 \times 10^{18} \pcc 
\int {\rm T}_{\rm B} \rm{(H~I~dv)}$/(K \kms). }
\end{figure}

To help distinguish between foreground and background material, or atomic
and molecular gases, we employed the H I profiles from the Leiden-Dwingeloo
all-sky H I survey \citep{HarBur97}. These data have 35\arcmin\ resolution 
on a 0.5\degr\ grid in galactic coordinates.

\subsection{Reddening and extinction}

The reddening toward \zoph\ is known to be \EBV\ = 0.32 mag \citep{Mor75}, 
but maps of the 
foreground extinction over only the first 140 pc are unavailable.  Instead, we 
employ the limiting reddening (from here to infinity) from the work of 
\cite{SchFin+98}, denoted by \EBVl, with a spatial resolution of 
6.1\arcmin, published on a 2.5\arcmin\ grid;  the stated global rms error of 
this dataset is 16\% (a percentage at each pixel).  Their values are
based on a determination of the dust column density estimated from the IRAS
100 micron flux adopting the temperature variation derived from
COBE/DIRBE 240 micron data.  The minimum limiting reddening 
in the region, approximately 
0.23 mag (Fig. 1), is likely hosted in atomic gas over long paths, as 
opposed to the more-localized diffuse and dark clouds of interest here. 

A comparison between the limiting reddening of \cite{SchFin+98} and
maps of the limiting extinction \AVl\ computed from star counts by
\cite{DobUeh+05} can be found in Sect. 7 (online only). 

\subsection{Some general conditions  in the gas along the line of sight to
\zoph}

In front of the star, \EBV\ = 0.32 mag,
N(H I) $ = 5.2 \times 10^{20}\pcc$,
N(\HH) $ =  4.5 \times 10^{20}\pcc$ \citep{SavDra+77},
N(\cotw) $\approx 2.4 \times 10^{15}\pcc$ \citep{Mor75,WanPen+82,SonWel+07},
 N(C$^0$) = N(\cotw) \citep{Mor75} and  
N(C\p) $\ga 3 \times 10^{17}\pcc$ \citep{CarMat+93}. 
The mean kinetic temperature of the molecular gas inferred from 
\HH\ absorption is 54 K \citep{SavDra+77}. 

Toward the star the limiting extinction from the work of 
\cite{SchFin+98} is \EBVl\ = 0.55 mag, so that the background reddening 
is approximately 0.23 mag.  

The distance to the occulting material is generally taken to be very 
close to that of the star, just outside the nearer edge of the star's
H II region \citep{WooHaf+05}.  
At a distance of 140 pc, 1\arcmin\ corresponds to 0.041 pc 
and 1\degr\ to 2.44 pc.

\section{The sky around \zoph\ viewed in reddening and CO emission}

Figure 1 is a composite image of the limiting reddening \EBVl\ and 
integrated \cotw\ J=1-0  emission, an updated version of Fig. 1 in 
\cite{TacAbe+00}.
The gray-scale underlay is the limiting reddening \citep{SchFin+98}
normalized to white at the minimum value seen over the region, 
\EBVl = 0.23 mag.  Superposed on the reddening map, the integrated 
\cotw\ intensity \WCO\ has been calculated separately over the 
diffuse/translucent northwest region, referred to as L121,
and the translucent/dark southeast region, L204.  For the diffuse
gas of L121 at upper right, the red and blue contours correspond 
to the velocity ranges above and below v = -0.25 \kms\ as indicated 
in the upper right corner: this division
corresponds to the natural separation between the two components
of CO, CH and OH emission found toward and around \zoph, as
shown in the profiles and position-velocity maps of \cite{Lis97}.  
For the darker gas of L204 to the southeast \WCO\ was calculated
over three intervals but nearly all of the emission from L204 occurs
in the velocity interval 2-6 \kms\ represented by the white
contours (see Fig. 5).   The noise level is somewhat larger for these contours,
because \WCO\ was calculated over a somewhat broader interval.  

The strong CO emission associated with L204 often follows the ridge 
lines of the extinction with something of a setback (reddening without
apparent CO) in the direction of the star.  This is consistent
with the edge-on geometry for L204 described by \cite{TacAbe+00}.
Nearly all of the CO emission in the L204 dark cloud complex
is found in the interval \vlsr\ = 2-6 \kms\ and so
does not overlap that of the more diffuse gas to the
north, seen at \vlsr\ $\la$ 2 \kms.  The extent to which the two regions 
are at rest in the directions joining them, and might partake of the optical
pumping excitation mechanism for diffuse gas described by 
\cite{WanPen+97} is unknown.  \cite{TacAbe+00} assumed that the
two clouds were co-moving at the edge of the H II region around
\zoph\ in order to discuss the energetics of the gas.

To the North of L204, on either side of the star is the \zoph\ 
diffuse cloud, described as the L121 complex by \cite{TacAbe+00}.   
Most of the emission from that gas occurs at 
-2.4 \kms\ $< $ \vlsr\ $ <$ 2.2 \kms.  The red-shifted
kinematic component at  -0.2 \kms\ $< $ \vlsr\ $ <$ 2.2 \kms which is seen
in CO emission and in many species in absorption toward the star, 
is very much confined to the northern and northwestern edges of 
the broader distribution of  blue-shifted gas.  Its separate identity 
as a second cloud is somewhat marginal in the emission maps and hardly
supported by a more detailed examination of the turbulent
gas kinematics (also see Fig. 7).

The diffuse gas in the L121 complex around \zoph\ is separated 
from the L204 dark cloud by an extended trough in the 
reddening whose overall mean value is \EBVl\ $\approx$ 0.43 mag. 
Immediately below the star is a pronounced minimum whose mean
is \EBVl\ $\approx$ 0.34 mag and whose absolute minimum is \EBVl\ =
0.29 mag.  The overall impression is of a cylindrical shell geometry
and perhaps a separate, more circular shell of radius $\approx$ 1\degr\  
around the star.  However, a yet-larger map of the extinction
shows that L204 is part of a much larger ridge.

\begin{figure}
\includegraphics[height=9cm]{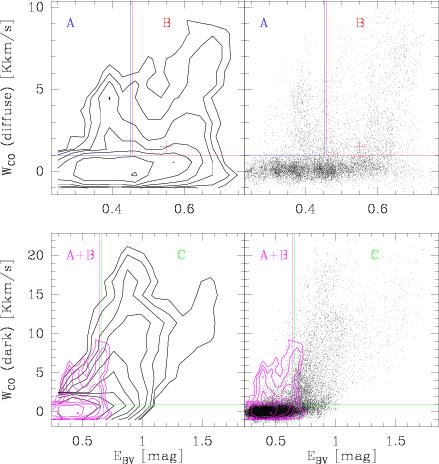}
\caption[]{Integrated intensity of the CO line \WCO\ {\it vs.}
limiting reddening \EBVl.  Top: for the diffuse region L121 at upper
right in Fig. 1; bottom: for the darker L204 gas at lower left in 
Fig. 1.  In each case the data are
shown twice, at right as individual data points, at left binned 
and contoured at logarithmic (factor two) intervals: 
the contour representation of the L121 data is superposed 
at bottom.  Also, in
each case the data are divided into regions of lower and higher
extinction with the separation occuring at the mean
\EBVl\ over datapoints which lack statistically significant
detections of CO emission (0.46 mag and 0.67 mag at
top and bottom, respectively).  The locus of the sightline toward 
\zoph\ in the B-region of the upper panel is indicated  by a 
(red) cross.  Mean \cotw\ emission profiles are shown in Fig. 5. }
\end{figure}

\subsection{Separating foreground and background reddening}

As noted above, we wish to understand the relationships between \WCO\
and reddening, but we only have maps of the reddening which include 
all of the unrelated background and foreground gas.    Immediately 
around \zoph\  \EBVl\ = 0.55 mag, compared to \EBV\ = 0.32 mag in 
front of the star, so some 0.55-0.32 mag = 0.23 mag of reddening 
occurs behind the star.  This also corresponds well to the 
absolute minimum in Fig. 1, \EBVl\ = 0.23 mag found some 1.5\degr\ 
North of the star.

To estimate the amount of extraneous foreground material, we note that 
this is expected to be in atomic form and, in absorption 
against the star, N(H I) $= 5.25 \times 10^{20}$ 
\citep{SavDra+77} corresponding to \EBV\ $\la$ 0.1 mag.  This 
leaves \EBV\ $\ga$  0.22 mag of foreground material associated 
with the pure \HH\ component and gives the general idea that 
perhaps as much as 0.32 mag should be subtracted from the 
map of \EBVl\ to infer the local reddening intrinsic to the \HH-bearing
gas.
However, this is probably an overestimate of the required correction 
because some of the foreground atomic gas is in the 
vicinity of L121 and L204, providing shielding.  

Figure 2 is a map 
of the integrated H I brightness, scaled to the mean relationship
between hydrogen column density and extinction, that is, the map shows 
$1.823 \times 10^{18} \pcc \int {\rm T}_{\rm B} \rm{(H I) dv}/
5.8 \times 10^{21} \pcc {\rm mag}^{-1}$ where the velocity integral
is in units of K \kms. Several of the local
reddening peaks which lack CO emission in Fig. 1 even at 
\EBVl\ = 0.65 mag are present as peaks in Fig. 2, for instance at 
(l,b) = (7.5\degr,21.5\degr) and (4\degr,20\degr).  Some of this 
gas must be indigenous to the region of interest, but
the extinction associated with the H I peaks is only $\approx 0.06$
mag, judging from the peak levels in Fig. 2 (0.28 mag) compared to the
nearby background level (0.22 mag).

The minima in Fig. 1 and Fig. 2 and the difference between the 
foreground and limiting reddening toward the star consistently 
imply a background reddening contribution of $\approx 0.2$ mag 
over the region of interest.

\subsection{Quantitative relationship between CO emission and reddening}

To quantify the relationship between reddening and CO emission, and
to compare and contrast the diffuse and dark sightlines we divided 
the extent of Fig. 1 along a Northeast-Southwest diagonal in the 
trough of reddening between L121 and L204, along a line running from 
(l,b) = (3\degr,19.333\degr) to (10\degr,25\degr). Fig. 3 shows
the integrated CO intensities \WCO\ and limiting reddening \EBVl\
for all points in both regions: the profile integral was taken 
over the range -2.4 \kms\ $<$ \vlsr\ $<$ 2.2 \kms\ for the diffuse 
L121 gas  shown in the top panel, and 2 \kms $<$ \vlsr\ $<$ 8 \kms\ for 
the darker L204 region in the lower portion of the map of Fig. 1 and 
the lower panel in Fig. 3.  The rms noise in integrated intensity 
is 0.5-0.6 K \kms.  Profile integrals above 1 K \kms\ generally 
represent real detections.

Once we realized that the CO emission in the diffuse region was
bimodal, as illustrated in the top panels of Fig. 3, we further 
sub-divided the diffuse gas into A and B portions corresponding to
the two branches of the emission distribution.  The A and B portions
were separated at \EBVl\ = 0.455 mag, which is the mean \EBVl\
for  those sightlines along which \WCO\ $<$ 1 K \kms and which 
therefore lack statistically significant detections of CO emission. 
The A-branch pixels have strong CO emission at \EBVl\ substantially 
below the mean of those sightlines lacking CO emission at all.
For the diffuse gas in L121 there is actually a substantial spatial 
segregation of the A and B portion pixels, with unweighted mean
$<($l,b$)>$ = (6.0\degr$\pm0.7$\degr,23.8\degr$\pm0.7$\degr) 
and $<($l,b$)>$ = (4.0\degr$\pm0.9$\degr,22.3\degr$\pm1.0$\degr)
for the A and B portions, respectively.  These centroids are on
opposite sides of \zoph\ and separated by more than
 $1\sigma + 1\sigma$ in each coordinate.   The red-shifted
gas appears mostly in the higher-extinction B-portion while
the blue-shifted gas appears more nearly 
in both the A and B-portions.

\begin{figure}
\includegraphics[height=9cm]{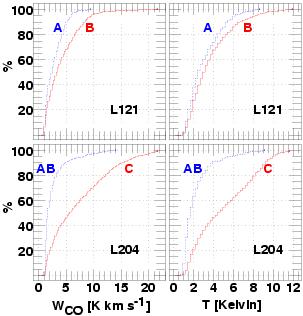}
\caption[]{Cumulative probability distributions of the 
integrated intensity \WCO\ (left) and peak line temperature (right) over
the diffuse, northeastern L121 (at top) and dark southeast L204 sightlines 
shown in  Fig. 1. Only those sightlines at which 
\WCO\ $\ge 1$ K \kms\ have been considered. }
\end{figure}

In either emission branch, \WCO\ varies widely over a relatively narrow
range in \EBVl, mimicing the extreme sensitivities of 
N(CO) which are seen in global absorption line surveys of single
sightlines over widely-separated regions (see \cite{Lis07CO}).
The widths of the two branches are comparable to the stated rms
noise of the \EBVl\ data (16\%) and they
are separated by 0.2 mag, which is of the same order as the
background contribution: it is possible that they would
more nearly coincide in \EBV\ if the background contribution were 
strongly
variable.  However, the B-branch at higher \EBVl\ is more heavily
 populated at \WCO\ $>$ 5 K \kms, suggesting that it actually is 
somewhat more strongly shielded, fostering a higher CO abundance
and brightness: in this regime, \WCO\ $\propto$ N(CO) ($ibid$ and 
see Sect. 6) and the conversion of free carbon from C\p\ to CO 
occurs over a very narrow interval in \EBV\ and/or N(\HH). 

The distribution of brightness in the darker L204 region is not
similarly bimodal at all \EBVl\ and there is less rationale for 
a simple division into sub-portions.  However when this 
is done, at \EBVl\ = 0.67 mag corresponding again to the mean over 
pixels lacking statistically significant CO emission, the resultant 
lower-\EBVl\ portion corresponds (in \EBVl\ and \WCO) to the 
entirety of the diffuse region.  Consequently the lower-\EBVl\ part
of L204 is labelled L204 AB and the other L204 C.  
The behavior of \WCO\ with \EBVl\ in 
the dark gas is complex but clearly bimodal for \WCO\ $\ga $ 12 K \kms;
the strongest emission is by no means limited to the darkest regions.
Statistics of the brightness distribution over L121 and L204
are shown in Fig. 4 and discussed in the following sections.

\subsection{Statistics of the line brightness}

Figure 4 presents the distribution of integrated and peak brightness.
The differences between the A (lower \EBVl) and B portions
of the diffuse gas are somewhat more pronounced in the distribution
 of the line profile integral at left and somewhat less so in the peak, 
so that the line widths differ more than the line heights.
Although the median brightness is at most only 3 K or 3 K \kms
in the diffuse gas and the tail of the distribution 
seems very poorly populated above, say, 8 K peak brightness: 
paths which traverse L121 even in the short dimension 
have a very substantial chance of containing at least 
one such bright line, as discussed in Sect. 5 (see Fig. 7).

It is not the intention here to discuss the dark gas, but it
should be noted that the difference in mean brightness 
between the diffuse and dark regions are modest and correspond 
approximately to the differences in \EBVl, thereby preserving the 
possibility of a common CO-\HH\ conversion factor; the same
is also true for the A and B portions of L121, see Fig. 5.  
This occurs despite 
the fact that most of the free gas-phase carbon is in C\p in L121 
(99\% toward \zoph) and in CO in the dark gas of L204 (where
N(CO) $\approx 3\times 10^{17} \pcc$) implying
a difference in CO column density and \WCO/N(CO) by a factor
of order 50.  In Section 5 we discuss the very different 
proportionality \WCO\ $\propto$ N(CO) which is observed 
 within the diffuse regime alone \citep{Lis07CO}.

\supjerome{For the darker gas of L204 in the bottom panels of Fig .3,
little CO is detected at reddening below 0.5 mag; material
corresponding to the lower-reddening branch of the L121 gas 
is either missing or indistinct.  At higher \EBVl $> 0.5$ mag,
\cotw\ emission seems to show three distinct branches and
it clearly bimodal at \WCO\ $>$ 12 K \kms. The higher-reddening
branch of the emission in L121 seems well-represented in
the lower panel of Fig. 3 where the increase of \WCO\ with 
reddening is rapid if perhaps not as steep as in 
the more diffuse regions.  The distribution of \WCO\ extends
to higher values in L204, but, conspicuously, at \EBVl\ $<$
1 mag, even though substantially larger \EBVl\ occurs
over the region.  The carbon in L204 probably resides mostly in 
CO and the higher values of \WCO\ probably reflect local
heating of the gas.}

\subsection{An incidental bound on N(\HH) over the diffuse gas}

  Because the line of sight to \zoph\ occurs 
at such a high value of \EBVl\ relative to the rest of L121 we infer 
that N(\HH) is never very much larger in L121 than toward the star.  
For instance, if we take \EBVl\ = 0.65 mag characteristic of the B-portion
and subtract a background contribution 0.23 mag equal to that toward
the star, the remaining gas column with \EBVl\ = 0.65 - 0.23 = 0.42 mag
corresponds to N(H) $\approx 2.4\times10^{21}$ H-nuclei $\pcc$ and
 N(\HH) $< 1.2 \times 10^{21}\pcc$.  Conversely, because the line of
sight toward the star has such a high value \EBVl\ = 0.55 mag relative
to the rest of the region, emission from the A-branch at \EBVl\ = 0.4 mag,
probably arises in regions whose reddening and N(\HH) are actually below 
those seen toward the star.

\begin{figure}
\includegraphics[height=9cm]{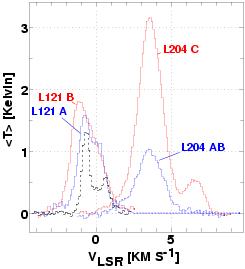}
\caption[]{Mean spectra over the diffuse and dark sub-portions 
defined in Fig. 3: the profiles in the diffuse gas (L121) 
are those at 
lower velocity. The integrated brightnesses are \WCO\ = 2.9 and
4.0 K \kms\ for L121 and 2.3 and 7.4 K \kms\ for L204.
Also shown (dashed black line) is a spectrum at 1\arcmin\ resolution 
toward \zoph\ from the 12m telescope. 
}
\end{figure}

\section{Line profiles, linewidths and turbulent flows in diffuse gas}

\subsection{CO line profiles and profile widths}

For a pure-\HH\ gas, the Doppler temperature equivalent to a given 
linewidth is T$_{\rm dopp} = 43.6$ K FWHM$^2$ or  
T$_{\rm dopp} = 242$ K \sv$^2$ where \sv\ is the velocity disperson 
in \kms. Typical kinetic temperatures in  CO-bearing diffuse gas 
are above 30 K 
\citep{SonWel+07,BurFra+07,Lis07CO} and the mean kinetic 
temperature seen in surveys of \HH\ absorption is 70 - 80 K 
\citep{RacSno+02,SavDra+77}.  CO profiles with FWHM $\la$
1 \kms\ are subsonic at diffuse cloud temperatures 
{\bf and the purely-thermal velocity dispersions of CO molecules, 
0.1 \kms\ at \TK = 50 K, do not contribute importantly to the
 observed linewidths}.

Unweighted mean profiles formed over the sub-portions of 
the diffuse L121 and dark L204 regions are shown in Fig. 5.  
The mean profiles have linewidths which are supersonic, FWHM of 
typically 2-3 \kms\ but the individual sightlines in L121 typically 
have subsonic linewidths; the FWHM of the two kinematic components 
seen toward the star at 1\arcmin\ resolution (shown in Fig. 5
as the dark dashed line) are 0.6 \kms\ and 1.1 \kms,  equivalent 
to Doppler temperatures of only 16 K and 53 K in a pure \HH\ gas,
compared to a measured temperature in \HH\ of 54 K as noted in Sect. 2.4.

The distribution
of velocity dispersions found over the L121 region is shown in
Fig. 6.  To produce this figure we used the following windowing
technique to measure the widths of spectra which might contain
more than one kinematic component: select the overall velocity 
interval of the diffuse gas;
find the peak channel; select those contiguous channels around this
peak with temperatures above a noise threshold of 0.25 K; calculate
the brightness-weighted velocity dispersion over those channels; mask 
off that portion of the profile; repeat until no channel above 1 K 
remains unmasked.  The dispersions measured {\it en masse} in this way 
are subject to overestimating the width in cases of unrecognizd blending, 
but they agree to within 10\% for a series of test profiles which were 
fit with gaussian components (for instance, those shown in Fig. 8).

The general properties of the gas in L121 have often been inferred from 
profiles seen along the single, microscopic absorption line of sight toward 
the star and  a sensitive CO profile toward \zoph\ at 1\arcmin\ spatial 
resolution is also included in Fig. 5.  How representative is it of the 
larger-scale distributions of the host gas?  The stronger CO component 
at \vlsr\ = -0.7 \kms\ toward \zoph\ is one of the narrowest lines
known in a diffuse cloud, with FWHM = 0.60 \kms.  Broad consideration 
of  this question is given in the discussion of Fig. 7 which displays most of the 
profiles seen in the L121 region, immediately following.

\subsection{Velocity gradients and turbulent flows}

 The connections between the profiles seen at
individual pixels and the mean profiles shown in Fig. 5 are the flows
and velocity gradients in the host medium, $i.e.$ the character of
the turbulence.  Although very elaborate analyses of 
line profile centroids \citep{PetFal03} and fluctuations in the wings
of optically thick profiles \citep{FalPhi96} have been used to infer
the properties of turbulence in denser neutral media, turbulent 
flows in the nearby diffuse gas of L121 are immediately visible 
in the line cores and (usually) spatially resolved into the 
shifts of individual, subsonic profiles.

\begin{figure}
\includegraphics[height=9cm]{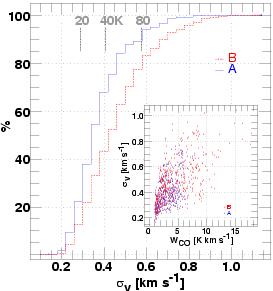}
\caption[]{Statistics of line profile velocity dispersions in the
diffuse gas. Large outer panel: cumulative probability distribution 
histograms of the measured velocity dispersion \sv\ shown separately 
for the A and B portions of the diffuse gas (see Sect. 4 of the 
text).  The {\bf one-dimensional} thermal velocity dispersions 
corresponding to kinetic temperatures of 20, 40 and 80 K in a 
gas of pure \HH\ are shown at upper left.
Inset at lower right: variation of \sv\ with \WCO.}
\end{figure}

This is illustrated for \zoph\ in Fig. 7, where we show a comprehensive
series of longitude-velocity diagrams spaced at one-pixel (4\arcmin) 
intervals in galactic latitude, as labelled in the individual panels.
The peak line brightnesses (K) seen in the individual panels of Fig. 7 
are labelled on the bars showing the color scaling in each panel.
Although lines with peak brightnesses above 8 K seem relatively 
rare in Fig. 4, they are common enough that peak brightnesses 7.5 K
and higher appear in 40\%, 18 of the 44 individual panels of Fig. 7.  
As shown 
in Fig. 3, the most intense lines in the diffuse gas are by no means 
limited to the more heavily-reddened sightlines.

Figure 7 shows that a description of the gas  in terms of two 
identifiable foreground clouds at the absorption line or CO emission 
line velocities toward \zoph\ is not appropriate.  The line of sight 
to the star could just as well have occured behind any of the 
profiles exhibited in Fig. 7, leading to a wide variety of possible
interpretations of the intervening medium.  Discussing the map in 
Fig. 1, we saw that the red-shifted component generally appears 
as something of a fringe at the northern edges of more 
broadly-distributed, negative-velocity gas.  Viewed in 
position-velocity space in 
Fig. 7, the red-shifted gas is often seen as a pronounced kinematic 
excursion or wing, for instance, at b = 22.8\degr\ or 
23.4\degr.  Further to the North, at b $> 23.2$\degr, the 
kinematic pattern undulates across the positive-velocity portion.
At b= 23.4\degr\ the resolved velocity gradient spans
the entire range of velocities in the diffuse gas: viewed from
a different perspective, this same region might have been seen as
a single, broad line.  

Figure 8 is an expanded view of the panel in Fig. 7 at 
b= 22.6\degr, where we also show several included line profiles,
their gaussian decomposition and the resultant FWHM.  Except
at the center of the diagram where there is a partially spatially-resolved
velocity gradient, the FWHM are small enough to be subsonic for an 
\HH\ gas at typical
diffuse cloud temperatures above 30 K.  Thus the observations show 
in detail how supersonic profiles might arise from the 
coincidental superposition or addition (for instance through
beam-smearing) of velocity-shifted quiescent regions.
   
As in Fig. 8 there are many highly-localized, 
relatively broad lines in 4\arcmin\ pixels, often joined to abrupt 
but spatially-resolved velocity gradients and reversals in 
narrow-lined gas.  This suggests that broader lines are themselves 
composed of unresolved velocity gradients and it seems possible
that any profile in Fig. 7 with a width substantially above sonic
is an unresolved gradient.  For instance, compare the velocity span at
b=22.87\degr, which is spatially resolved, with that at the edge
of the emission region at b=23.33\degr.  Recalling the extremely
narrow blue-shifted CO component toward \zoph\ and the low 
implied Doppler temperature, we wonder whether and at what scale 
profiles having linewidths that are subsonic at the higher
temperatures expected for diffuse gas, say 60 K, might show
velocity or spatial substructure.   The geometry of the turbulent
flows producing the patterns in Fig. 7 will be considered in
a subsequent paper.

\begin{figure*}
\includegraphics[height=19.4cm]{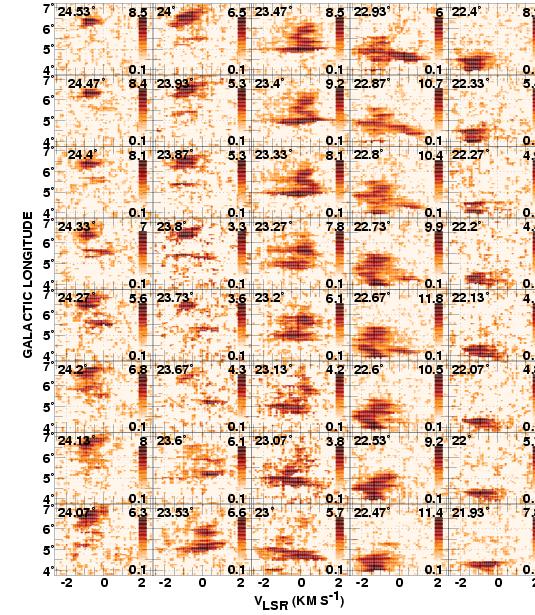}
\caption[]{Longitude-velocity diagrams traversing the \zoph\ diffuse 
cloud at many latitudes within the Northwest portion of the region 
shown in Fig. 1. The vertical separation between panels is 4\arcmin\
(1 pixel) and the galactic latitude within each panel is labelled. The 
color scale in each panel are separately normalized 
within a range (Kelvins) shown on the individual color bars.}
\end{figure*}

\section{Physical conditions in the diffuse gas}
  

There are some 1000 sightlines with \WCO\ $\ga$ 1 K \kms\
in the diffuse, Northwest portion of the map in Fig. 1.  At a distance
of 140 pc, the \HH-mass associated with these sightlines, parametrized
by their average \HH\ column density $<$N(\HH)$>$, 
is M = 430 $\Msun <$ N(\HH)/$1\times 10^{21}\pcc>$  where
typical values of N(\HH) along the L121 sightlines are 
$0.5 - 1.2 \times 10^{21}\pcc$ (see Sect. 2.4 and 3.4). This may be compared
with the value  520 $\Msun$ given by \cite{TacAbe+00} based on a
CO-\HH\ conversion factor N(\HH)/\WCO\ = $1.56\times10^{20}\pcc$/(K \kms).

Nonetheless, deriving the physical parameters of host diffuse gas 
from CO profiles is challenging.  Strong fractionation of the carbon 
isotopes \citep{LisLuc98,SonWel+07,BurFra+07,Lis07CO} causes the 
N(\cotw)/N(\coth) to vary in the range 15 - 150, thus making it 
impossible to derive the line excitation temperatures and optical
depths, or the kinetic temperatures and column densities, under 
the usual assumption (valid in dark gas) that the relative abundances of 
\cotw\ and \coth\ only reflect the general interstellar carbon isotope ratio.
However, it is straightforward to show that the general properties
of the CO observations, even including the rather unexpectedly bright
lines, fit easily into the framework of diffuse gas at typical
temperatures 30 - 60 K and modest density.  The real underlying 
mysteries are the working of the
poorly-understood polyatomic chemistry which forms the CO and other
species at such modest densities \citep{LisLuc+06}, and the origin
of the turbulent flows which may power the chemistry.

Figure 9 reports some results from the models which were used to interpret
CO absorption line data by \cite{Lis07CO}.  The underlying physics are:
to heat a uniform-density gas sphere by the photoelectric effect on small 
grains, as in the work of \cite{WolHol+95,WolMcK+03}, to calculate the 
ionization balance including grain-assisted neutralization of atomic 
ions (including most importantly the protons, $ibid$), to allow 
equilibrium of the self-shielding of \HH\ formed on grains and 
the formation of CO by thermal electron-recombination of a 
fixed quantity of \hcop\  X(\hcop)  = N(HCO\p)/N(\HH) = $2\times 10^{-9}$ 
 (the actual secular evolution is traced
by \cite{Lis07}); and, to calculate the rotational excitation of CO assuming
microturbulent radiative transfer with a linewidth determined by the
local sound speed.  Note that the interstellar radiation field in the
models has not been increased above the mean interstellar value 
to account for the presence of the star
and that all of the aspects shown here for the model results
are present in the absorption line data summarized by \cite{Lis07CO}.

Model results are shown for just two fairly high densities 
n(H) $= 256\pccc$ and $512 \pccc$.  To form the graphs in Fig. 9, 
results were derived by integrating along sightlines ranging across 
the faces of model 
spheres from center to edge. The graphs summarize results gleaned 
from models whose central column densities N(H) varied widely, so 
that the same value of N(\HH) might occur at different positions
across the faces of models with differing n(H) and N(H), and therefore
have different N(CO) and \WCO.  For N(\HH) = $1\times 10^{21}\pcc$ the models have typical 
sizes of 1.2-2.5  pc but the CO abundance is concentrated into 
smaller central portions of the host bodies owing to
the chemistry of CO and that of \HH, and the CO emission is more concentrated
still owing to geometry and radiative pumping.

As noted above N(\HH) $ = 0.5 - 1\times10^{21}\pcc$ in the CO-emitting
regions around the star.  As shown in panel c at the lower left, 
this is precisely the 
regime where carbon is about to recombine fully to CO at the quoted 
densities: the increase of N(CO) with N(\HH) is very rapid.  
Both N(CO) and \WCO\ vary rapidly and have large scatter when plotted
against N(\HH).  Substantial CO column densities can accumulate in gas 
which is still relatively warm, 30-50 K, giving rise to 12 K lines as 
observed in the  brightest profiles.  

The most important consideration is as shown at the upper right in 
panel b, \WCO\ $\propto$ N(CO)  for \WCO\ $\la$ 10 K \kms,  explicitly 
independent of density and implicitly independent of N(\HH) and
other cloud properties.  This is a very general consequence of
very sub-thermal excitation, as first shown by \cite{GolKwa74} 
and does not require low optical depth.
As shown in Fig. 4, some 80-90\% of the 
diffuse cloud spectra have \WCO\ $<$ 5-6 K \kms\ and virtually
all are below 10-12 K \kms,  just in the regime 
characteristic of sightlines studied in $uv$ and mm-wave CO absorption 
work generally, see Fig. 6 of \cite{Lis07CO}.  Therefore,
{\it the \cotw\ brightness map in Fig. 1 is a map of N(\cotw) in
the diffuse gas} and the same would be true for N(\coth) and 
the brightness of the \coth\ line.  This is the one unambiguous
result of mapping CO emission in any diffuse gas.

The extreme sensitivity of \WCO\ to N(\HH) therefore arises because 
\WCO\ $\propto$ N(CO), so that a plot of \WCO\ $vs.$ N(\HH) is 
equivalent to plotting N(CO) against N(\HH).  The net result is that
although commonly-used values of the CO-\HH\ conversion factor apply
to some gas (as indicated in panel a of Fig. 9) the CO-\HH\
conversion factor varies widely in diffuse gas and the actual 
N(CO)/N(\HH) ratio is small but very uncertain.    A map of 
CO emission from diffuse gas is an image of the {\it chemistry}, 
not the mass distribution.

Last, note that the J=1-0 CO line brightness is insensitive to density at 
fixed N(CO), indicating that other tracers are 
required to measure the local density when mm-wave emission profiles 
are analyzed.  The J=2-1/J=1-0 line brightness ratios at lower 
right in Fig. 9 are not very sensitve to density, which explains 
why  line brightness ratios 0.7 - 0.75 are indeed so commonly observed
in diffuse and translucent gas \citep{FalPan+98,PetLuc+08}.  The J=3-2/J=1-0
brightness ratio is a better indicator of density in the CO lines, but
care must be taken to match the spatial resolution of the two lines.

\begin{figure}
\includegraphics[height=9.8cm]{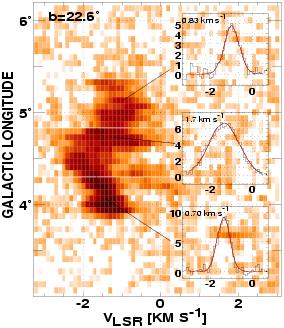}
\caption[]{The longitude-velocity diagram at b = 22.6\degr\ from Fig. 
7, expanded and decomposed into its constituent profiles at three 
positions (shown inset).  Overlaid on each inset spectrum is a one-component 
gaussian fit whose FWHM is shown at the upper left.}
\end{figure}

\section{Summary}

The line of sight to the nearby (140-160 pc) runaway 09.5V star 
\zoph\ has for many years been used as an archetype for studying 
the properties of diffuse clouds in optical and $uv$ absorption.  
Because the material has an appreciable molecular content, the host 
diffuse clouds  can actually be imaged on the sky in space 
and radial velocity.  Because the gas is well extended and comparatively 
close, it provides an unusual opportunity for study of diffuse gas 
and its interaction with its surroundings, including the star.

\begin{figure}
\includegraphics[height=8.5cm]{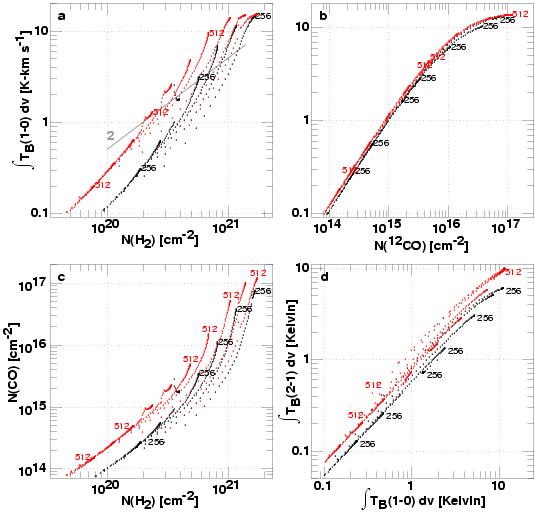}
\caption[]{Models of \cotw\ formation and excitation.  Results are shown for models
having n(H) = n(H I) + 2n(\HH) = $256\pccc$ and $512\pccc$.  Upper left: Variation
of integrated \cotw\ brightness \WCO\ with N(\HH); the shaded line labelled ``2''
is a CO-\HH\ conversion factor $2\times10^{20}~\HH \pcc$ (K \kms)$^{-1}$.  
Upper right: variation of \WCO\ with N(\cotw).
Bottom left: variation of N(\cotw) with N(\HH).  Lower right: comparison of
\cotw\ line brightnesses in the lowest two rotational transitions.}
\end{figure}

We began by comparing the \cotw\ J=1-0 emission line datacube of
\cite{TacAbe+00} (HPBW 2.7\arcmin\ observations on a 4\arcmin\ = 0.1 pc 
pixel grid)
with a map of the limiting reddening \EBVl\ from the work of \cite{SchFin+98}
having similar resolution 6.1\arcmin\ (Fig. 1).  The reddening in the L121
complex near and around \zoph\ ranges from \EBVl\ = 0.23-0.75 mag and 
the integrated CO emission up to \WCO\ = 12K \kms, with 12 K peak 
temperatures, which are very bright lines indeed.  
Comparison of reddening of the star (0.32 mag) and through the Milky 
Way (\EBVl = 0.55 mag) in the same direction, and 
comparison of H I seen toward the star in $uv$ absorption and around 
the star in 21cm emission (Fig. 2), suggests that $\approx 0.2$ mag of 
\EBVl\ should be ascribed to unrelated background material. 

CO emission from diffuse L121 gas seen around \zoph\ is bimodal 
in \EBVl, clustering around \EBVl\ = 
0.4 mag and 0.65 mag and varying widely (1 K \kms\ $<$ \WCO $<$ 6-12 K \kms) 
with \EBVl\ in one of two narrow ranges, see Fig. 3:  the same large 
scatter in CO column density with reddening and N(H) which is seen 
globally in galactic absorption line surveys also occurs in the 
single region studied here.  
The lower-reddening branch of the emission is spatially 
segregated to the galactic northeast of the star and has somewhat 
smaller mean integrated brightness \WCO\ (Fig. 4) and velocity dispersion \sv\ 
(0.35 \kms\ vs. 0.42 \kms; Fig. 6).  However, peak temperatures 
8-12 K are present in both branches.   

The two most striking observational results of this study are the 
strong lines which emanate from the diffuse gas, up to \WCO\ = 12 K \kms, 
and the velocity structure present in the strongly emitting \cotw\ line 
cores there (Figs. 7 and 8). 
The turbulent flows in this gas are in general directly visible as the 
spatially and kinematically resolved velocity gradients in simple, 
narrow, bright lines whose widths (\sv\ $<$ 0.6 \kms) are subsonic at 
diffuse cloud temperatures  \TK\ = 30 - 60 K (Fig 4 and 8).  Other, 
locally-broader line profiles will likely be resolved into such
velocity shifts of narrow line cores with higher 
(than 4\arcmin) resolution although this remains to be tested.
Conversely, it also remains to be seen whether CO line profiles which
are subsonic but still supra-thermal at 2.7\arcmin\ resolution
will show spatial or velocity sub-structure when mapped at higher
resolution.  

We briefly discussed some modelling results of the formation
and excitation of CO in diffuse media (Fig. 9).  The brightness 
of the strongest CO lines can be understood by noting that the \HH\ 
column densities in the gas around \zoph\ are near the point 
(N(\HH) $\approx 10^{21}\pcc$) where carbon fully recombines to CO
at even modest densities n(H) $= 200-500\pccc$, so that substantial
columns of CO may form in gas which is at typical diffuse cloud
temperatures (above 30 K).  In turn, such densities will excite CO
to the required degree even though they are far too low to 
thermalize the lower rotational level populations.  

In the range  \WCO\ $\la$ 10 K \kms\ 
it is the case that \WCO\ $\propto$ N(CO), as a consequence
of very sub-thermal excitation.  The CO-\HH\  
conversion factor therefore varies widely in diffuse gas 
(because N(CO) varies rapidly with N(\HH) and with great scatter), 
but it takes on values N(\HH)/\WCO\ $=2  \times 10^{20} \HH \pcc$/(K \kms)
in limited circumstances (Fig. 9 panel a at upper left) .  

If \WCO\ $\propto$ N(CO), a sky map of \WCO\ like that in Fig. 1 
is a map of the interstellar diffuse cloud chemistry.  This should be
contrasted with the more usual assumption of a constant CO conversion
factor  N(\HH)/\WCO, in which case a CO map traces the contours 
of the mass ($i.e.$ the bulk molecular material) 
largely independent of physical conditions.
Even the darker gas seen in the L204 complex South of \zoph\ is not
immune to the influences of chemistry, which are clearly visible 
in the displacements between regions of higher \WCO\ and \EBVl. 

The wealth of structure seen in the foreground CO brightness map 
has important implications for absorption line study of diffuse clouds.   
 Since the viewing geometry is an accident of the relative locations
of the Sun and \zoph, our line of sight to the star could equally
well have intercepted any of the wide variety of profiles shown in 
Fig. 7.  This belies our ability to infer the general properties of 
the host gas from studies along any single sightline, no matter 
how comprehensive;  studies of individual absorption sightlines must be viewed 
demographically, as datapoints within a large family of possible 
outcomes, even in nominally similar conditions.
Even beyond this, there are some obvious fundamental limits to the
use of absorption lines to derive the properties of the intervening
gas.   In the present case only a map could
correct the false impression that the star is occulted 
by two separate foreground clouds.   Likewise, the turbulent
flows in the foreground gas appear clearly in maps of the gas
but not at most individual positions, thereby conveying the false
impression of an overly-quiescent medium.

We intend to map with higher angular resolution some regions of 
the L121 gas whose line profiles are thermal at the 
2.7\arcmin\ NANTEN resolution, to see what kind of substructure 
might be present when some forms of line-broadening are absent.  
We will also map some L121 gas whose profiles are broader and whose 
velocity gradients are not fully resolved in Fig. 7 at 
4\arcmin\ (0.16 pc) beam-spacing.  

This is the second paper in a loose series (see \cite{PetLuc+08}) which
will also report observations of similar kinematics on smaller 
angular scales and at higher angular resolution 
6\arcsec - 22\arcsec\ in other diffuse clouds of unknown distance 
whose presence was first manifested in our mm-wave absorption 
studies of polyatomic molecules. 
In a subsequent paper we will discuss the geometrical and physical
interpretation of the internal structures responsible for the flows 
seen there and in Fig. 7. 

\begin{acknowledgements}

The National Radio Astronomy Observatory is operated by Associated 
Universites, Inc. under a cooperative agreement with the US National 
Science Foundation.   IRAM is
operated by CNRS (France), the MPG (Germany) and the IGN (Spain).  
This research made use of the Simbad astronomical database and the
NASA ADS astrophysical database system.  This work profited from
 discussions of CO excitation with Michel Guelin.

\end{acknowledgements}
 
\bibliographystyle{apj}

\Online{}

\section{Comparison of measured \EBVl\ and \AVl}

As an alternative source of extinction data we considered the results
of \cite{DobUeh+05} who constructed a sky map of limiting extinction 
\AVl\ at latitudes $|b| < 40$\degr\ based on star counts. We generated
\AVl\ from a FITS file downloaded from the survey website, with pixels on
a 2\arcmin\ grid. The features in a map of that data strongly resemble 
those shown in Fig. 1 but \AVl\ = 0.25 mag toward \zoph, which substantially 
underestimates the known foreground extinction since it is accepted that 
\EBV\ = 0.32 mag and \AV\ $\approx 3.1 \times 0.32 = 1.0$ mag.

A more general comparison with the limiting extinction of \cite{SchFin+98} 
is shown in Fig. 10:  we generated \AVl\ and \EBVl\ on a 4\arcmin\ grid 
for 3\degr\ $ \le l \le$ 9\degr, 18\degr\ $ \le b \le$ 25\degr\ as in 
Fig. 1.   The \EBVl-intercept (0.32 mag) and slope (0.41) indicate 
an offset of 0.32 mag/0.41 = 0.78 mag in the \AVl\ dataset with respect
to \EBVl, which corresponds to the disparity in \EBVl\ and \AVl\
values toward the star itself or to the minimum \EBVl\ = 0.23 mag over 
the region.  We infer that a uniform foreground component of the 
extinction, which might have renormalized the star count, is absent
 in the \AVl\ maps for this region.
 
\begin{figure}
\includegraphics[height=8.7cm]{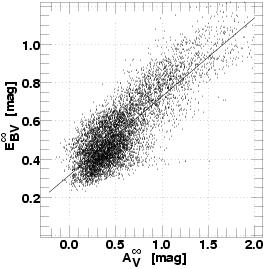}
\caption[]{Comparison of the limiting extinction \AVl\ of \cite{DobUeh+05}
with the limiting reddening data \EBVl\ of \cite{SchFin+98}. There is a
left-offset of the \AVl\ data of 0.78 mag, which is also seen toward
\zoph\ itself, see Sect. 6 of the text (available online).}
\end{figure}



\begin{figure*}
\includegraphics[height=19cm,angle=-90]{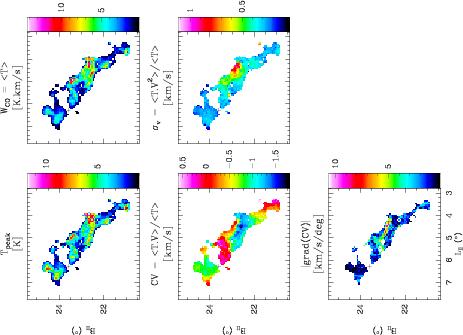}
\caption[]{Maps of various quantities,  available online.
Top left and right,
peak and integrated brightness.  Middle: mean velocity and 
velocity dispersion.  Bottom, the velocity gradient..}
\end{figure*}



\end{document}